\documentclass[aps,pra,twocolumn]{revtex4-2}
\usepackage{color}
\usepackage{longtable}
\usepackage{dcolumn}
\usepackage{graphicx}
\usepackage{bm}
\usepackage{bbm}
\usepackage{times}
\usepackage{nicefrac}
\usepackage{amsmath}
\usepackage{amsfonts}
\usepackage{amssymb}
\usepackage{amsthm}

\usepackage{physics}

\newcommand{\lbr}{\langle}
\newcommand{\rbr}{\rangle}

\newcommand{\Za}{Z\alpha}

\allowdisplaybreaks

\newcolumntype{w}[1]{D{.}{.}{#1}}

\begin{document}
\preprint{Version 1.0}

\title{Heavy-particle quantum electrodynamics}

\author{Krzysztof Pachucki}

\affiliation{Faculty of Physics, University of Warsaw,
             Pasteura 5, 02-093 Warsaw, Poland}

\author{Vladimir A. Yerokhin}
\affiliation{Max–Planck–Institut f\"ur Kernphysik, Saupfercheckweg 1, 69117 Heidelberg, Germany}

\date{\today}
\begin{abstract}
The quantum electrodynamic formalism is presented for the systematic and exact in $Z\,\alpha$ derivation
of nuclear recoil corrections in hydrogenic systems.

\end{abstract}

\pacs{31.30.jr, 36.10.Ee, 14.20.Dh}

\maketitle

\section{Introduction}
The quantum electrodynamics (QED) of atomic systems has so far been formulated for an infinitely heavy nucleus,
using the so called Furry picture, in which the Dirac propagator includes the Coulomb potential \cite{shabaevPR}.
The inclusion of finite nuclear mass effects in the exact relativistic formalism is highly nontrivial.
The unperturbative in the $Z\,\alpha$  formula for the leading recoil correction to the binding energy was originally
derived about 40 years ago by Shabaev in Refs. \cite{shabaev:85, shabaev:88}
and was confirmed by an independent derivation in Ref. \cite{pachucki:95}. This formula formed the basis of extensive numerical
QED calculations of nuclear recoil effects in heavy hydrogenic \cite{artemyev:95:pra, artemyev:95:jpb} and  heavy few-electron ions within the so-called
$1/Z$ expansion \cite{few_recoil}. This leading recoil correction was later extended to the presence of the homogenous magnetic field \cite{shabaev_01},
and recently to the finite-size nucleus \cite{fsrec},  but no other progress has been achieved so far.

Nuclear recoil effects are significant for light hydrogenic systems, where a different QED approach has been developed.
These corrections have been calculated over many years using $Z\,\alpha$ expansion, which led to very accurate results
and consequently the accurate determination of fundamental constants and quantum electrodynamics tests.
Nevertheless, some important higher-order recoil corrections are not yet known and limit theoretical predictions.
For example, the radiative recoil correction $\alpha\,(Z\,\alpha)^6\,m/M$ currently limits theoretical predictions for the hydrogen Lamb shift and
the muonium hyperfine structure \cite{codata}. This is due to the high complexity in calculations of higher order (in $\alpha$ and $Z\,\alpha)$
recoil corrections. Some of them have been calculated only by one group, such as the $(Z\,\alpha)^2\,m/M$ correction to
the hyperfine splitting in hydrogen about 40 years ago by Bodwin and Yennie  \cite{Bodwin:88}, which has not yet been confirmed.
Therefore, the exact in $Z\,\alpha$ formulas for recoil and radiative recoil corrections
of arbitrary order in mass ratio would be very desirable.

Very recently, the method from Refs. \cite{pachucki:95, fsrec}  has been extended to derive the exact, in $Z\,\alpha$, pure recoil correction
to the hyperfine splitting in hydrogenic systems \cite{hfs_rec}. It was noticed there, that the use of the temporal gauge for
the photon exchange propagators greatly simplifies the derivation and final formulas, so they
can be a basis for direct numerical calculations, but also they can be a basis for the analytic derivation of $Z\,\alpha$ expansion coefficients,
such as that of Bodwin and Yennie \cite{Bodwin:88}.

In this paper, we demonstrate that the method from Ref. \cite{hfs_rec} can be further developed to derive exact, in $Z\,\alpha$, formulas for
corrections of an arbitrary order in the mass ratio and $\alpha$. As an example, we present them for the leading radiative recoil
correction and for the complete non-radiative second order in mass ratio correction.
Finally, we perform an exemplary numerical calculation of nuclear recoil with the electron vacuum-polarization for muonic atoms.
Theoretical units $\hbar = c = 1$ are used throughout this work with $\alpha = e^2/(4\,\pi)$ and $e$ being an electron charge.

\section{Expansion in the electron-nuclear mass ratio}
To begin with the precise formulation of the nuclear recoil correction,
let us note that on the basis of  QED theory, the binding energy of a two-body system is a function of
$\alpha, Z\,\alpha$, and the mass ratio $m/M$, namely $E=E(m/M, Z\,\alpha,\alpha)$.
Assuming that one of the particles is much heavier and of extended size, one can perform an expansion in the mass ratio
together with the expansion in the fine structure constant $\alpha$,
 \begin{align}
 E\Big(\frac{m}{M},Z\,\alpha,\alpha\Big) =&\ \sum_{i,j} E^{(i,j)}(Z\,\alpha)\,, \label{01}
  \end{align}
  where  the dependence on the finite nuclear size, typically $m\,r_C$, is not explicitly shown, and where
 \begin{align}
 E^{(i,j)}(Z\,\alpha) =& m\,\Bigl(\frac{m}{M}\Bigr)^i\,\alpha^j\,{\cal E}^{(i,j)}(Z\,\alpha)\,. \label{02}
 \end{align}
The power of  $\alpha$ represents the number of QED loops, namely the number of the lepton self-energy and the vacuum-polarization loops.
Here,  $E^{(0,0)}(Z\,\alpha) = E_D$ is  the Dirac energy in the infinite nuclear mass limit,
\begin{align}
H_D\,\phi = E_D\,\phi\,, \label{03}
\end{align}
where
\begin{align}
H_D = \vec\alpha\cdot\vec p + \beta m +V_C\,, \label{04}
\end{align}
and where $V_C$ is a Coulomb potential  including the nuclear charge distribution $\rho_C(r)$,
\begin{align}
V_C(r) =&\ - \int d^3r'\,\frac{Z\,\alpha}{|\vec r-\vec r\,'|}\,\rho_C(r')\,. \label{05}
\end{align}
The next term, $E^{(0,1)}(Z\,\alpha) = E_\mathrm{self} + E_\mathrm{vp}$, is the sum of the
one-loop electron self-energy and vacuum-polarization corrections.
Namely, the electron self-energy correction is
\begin{align}
E_\mathrm{self} =&\ \langle \bar \phi|\Sigma_\mathrm{rad}(E_D) |\phi\rangle\,, \label{06}
\end{align}
where $\bar\phi = \phi^+\,\gamma^0$, and
\begin{align}
\Sigma_\mathrm{rad}(E) =&\
e^2\!\int\!\frac{d^4k}{(2\,\pi)^4\,i}\,\frac{1}{k^2}\,\gamma^\mu\,e^{-i\,\vec k\cdot\vec r}\,
S_F(E+\omega)\,\gamma_\mu\,e^{i\,\vec k\cdot\vec r} \label{07}
\end{align}
with $\omega = k^0$ the Feynman integration contour is assumed, and where
\begin{align}
S_F(E) =&\ \frac{1}{\not\!p-\gamma^0\,V_C-m} \label{08}
\end{align}
is the Dirac-Coulomb propagator.

In the second contribution $E_\mathrm{vp}$, the electron vacuum polarization modifies the photon propagator,
which is exchanged between the lepton and the nucleus
\begin{equation}
-\frac{g_{\mu\nu}}{k^2}\rightarrow - \frac{g_{\mu\nu}}{k^2(1+\bar{\omega}(k^2))}\,. \label{09}
\end{equation}
At the one--loop level $\bar{\omega}$ (in electron mass units) is given by
\begin{align}
\bar{\omega}(k^2) =&\ \frac{\alpha}{\pi}\,k^2\,
\int_4^\infty\,d(q^2)\frac{1}{q^2(m_e^2\,q^2-k^2)}\,u(q^2)\,, \label{10}
\end{align}
where
\begin{align}
u(q^2) =&\ \frac{1}{3}\sqrt{1-\frac{4}{q^2}}\,\left(1+\frac{2}{q^2}\right)\,. \label{11}
\end{align}
The resulting vacuum polarization potential for a point infinitely heavy nucleus is given  by
\begin{equation}
V_\mathrm{vp}(r)  =  -\frac{Z\,\alpha}{r}\,
\frac{\alpha}{\pi}\,\int_4^\infty \frac{d(q^2)}{q^2}\,e^{-m_e\,q\,r}\,u(q^2)\,,
\label{12}
\end{equation}
while for the finite size nucleus it is
\begin{align}
V_\mathrm{Cvp}(r) =&\  \int d^3r'\,V_\mathrm{vp}(|\vec r-\vec r\,'|)\,\rho_C(r')\,, \label{13}
\end{align}
and the corresponding vacuum polarization correction to the binding energy is
\begin{align}
E_\mathrm{vp} =&\ \langle \phi| V_\mathrm{Cvp} |\phi\rangle\,. \label{14}
\end{align}
These QED corrections in the infinite nuclear mass limit can be formally extended to an arbitrary number of loops,
and calculated numerically (see, for example, the two-loop self-energy calculation of the Lamb shift by one of us  \cite{twlp_num}).

\section{HPQED}
Heavy-particle quantum electrodynamics (HPQED) is the formalism that allows to derive nuclear recoil corrections of an arbitrary order in $m/M$ and $\alpha$.
The starting point is  the nonrelativistic QED Hamiltonian for the nucleus, namely
\begin{align}
H_\mathrm{nuc} =&\ \frac{\vec\Pi^2}{2\,M} + q\,A^0 -\frac{q}{2\,M}\,g\,\vec I\cdot\vec B - \frac{q\,\delta_I}{8\,M^2}\,\vec\nabla\cdot\vec E
\nonumber \\ &\
-\frac{q}{4\,M^2}\,(g-1)\,\vec I\cdot[\vec E\times\vec\Pi-\vec\Pi\times\vec E]+\ldots\,, \label{15}
\end{align}
where $\vec\Pi = \vec P-q\,\vec A$, $q = -Z\,e$, $I$ is the nuclear spin, $\delta_0 = 0, \delta_{1/2} = 1$,
and where we introduced the nuclear $g$ factor, defined as
\begin{align}
\vec\mu = \frac{q}{2\,M}\,g\,\vec I\,. \label{16}
\end{align}
We will neglect the nuclear quadrupole and all higher electromagnetic moments, because their contribution is usually very small.
Moreover, the finite charge and magnetic moment distributions, which are described by electromagnetic form factors
$G_E(-k^2)$ and $G_M(-k^2)$ (normalized to 1 at $k^2=0$) will be moved to the photon propagator. This can be done, because every photon exchange
between the point lepton and the nucleus is associated with the product of the nuclear vertex which contains
these form factors and of  the photon propagator.
Moreover,  we will assume at the beginning that all these form factors are the same, namely $G_E(-k^2) = G_M(-k^2)=\rho_C(-k^2)$.
This can be generalized at the final stage of derivation. We will also neglect nuclear polarizability, namely all  interactions beyond
the elastic form factors. Its calculation would require a separate treatment.
Moreover, the diagrams involving the photon emission and absorption
by the nucleus will also be neglected. Their consideration is very problematic when an elastic
approximation is assumed. This is because, even for light elements such as Mg $(Z=12)$, the effective electromagnetic coupling
$Z^2\,\alpha$ is larger than 1 and the QED perturbation theory may not work in such a case, 
which demonstrates limitations of the elastic form factor approximation.

Returning to the nuclear Hamiltonian in Eq. (\ref{15}) we construct the quantum electrodynamic theory
using the Coulomb gauge and the Furry picture. Namely, the Coulomb interaction  $V_C$ between the finite size nucleus
and the lepton is accounted for unperturbatively, and all other interactions are treated using the perturbation theory.
At the zeroth order, we have the standard QED with the static Coulomb potential, as described in the previous section.
The first-order term in the mass ratio is represented as an expectation value
\begin{align}
E^{(1)} =&\ \bigg\langle\Psi\bigg| \frac{(\vec P-q\,\vec A)^2}{2\,M} -\frac{q}{2\,M}\,g\,\vec I\cdot\vec B\bigg|\Psi\bigg\rangle_\mathrm{QED}
\label{17}
\end{align}
on a hydrogenic state $|\Psi\rangle$, which is a second-quantized Fock state centered at the position of nucleus $\vec R$.
The meaning of the``QED" expectation value needs to be explained and we here follow Refs. \cite{shabaevPR, hfs_rec}.
The matrix element of an arbitrary operator $Q$ on a state $\Psi$ in a QED theory is
\begin{align}
\langle\Psi|Q|\Psi\rangle_\mathrm{QED} =
\frac{\langle\Psi|{\mathrm T} Q\,\exp[-i\int d^4y\, H_I(y)]|\Psi\rangle}{\langle\Psi|{\mathrm T}\exp[-i\int d^4y\, H_I(y)]|\Psi\rangle}\,, \label{18}
\end{align}
where T denotes chronological ordering with an assumption that the time coordinate of $Q$ is $t=0$,
the interaction Hamiltonian is
\begin{align}
H_I(y) = e\,j_\mu(y) A^\mu(y)\,, \label{19}
\end{align}
and $j^\mu$ is the four-vector current.

The second term in Eq. (\ref{17}) leads to the well-known hyperfine splitting, and its calculation is standard.
The crucial point is the interpretation of $\vec P$ from the first term in Eq. (\ref{17}) and its action on $|\Psi\rangle$ and $\hat \psi$.
Namely, let us consider the representation of the fermion field $\hat \psi$
in terms of creation and annihilation operators of one-particle hydrogenic states $\phi_s$,
\begin{align}
\hat\psi(x)=&\ \sum_s^+ a_s\phi_s(\vec x)\,e^{-i\,E_s t} + \sum_s^- b_s\phi_s(\vec x)\,e^{-i\,E_s t}\,,
\nonumber \\
\hat\psi^+(x)=&\ \sum_s^+ a^+_s\phi^+_s(\vec x)\,e^{i\,E_s t} + \sum_s^- b^+_s\phi^+_s(\vec x)\,e^{i\,E_s t}\,. \label{20}
\end{align}
The differentiation $\vec\nabla_R$  acts on functions $\phi_s$ and
operators $a_s, b_s$, and this can be represented as \cite{hfs_rec}
\begin{align}
\vec \nabla_R =&\ \int d^3r\, \hat \psi^+(\vec r)\,\vec \partial_R\,\hat \psi(\vec r) + \vec\partial_R
\nonumber \\ =&\
-\int d^3r\, \hat \psi^+(\vec r)\,\vec \partial_r\,\hat \psi(\vec r) + \vec\partial_R \label{21}
\end{align}
where $\hat\psi(\vec r) \equiv \hat\psi(0,\vec r)$, and $\vec\partial_R$ is understood in the following sense.
The hydrogenic state $\phi_s$ is a function of $\phi_s(\vec r -\vec R)$ of the difference in electron  and nucleus position vectors;
therefore  $\vec \partial_R\,\phi_s = -\vec \partial_r\,\phi_s$, and $\hat a_s, \hat b_s$ remain intact.
As a test, for $t=0$,
\begin{align}
\vec\nabla_R\hat\psi(0, \vec x) =&\ -\int d^3r\, \hat \psi^+(\vec r)\,\vec \partial_r\,\hat \psi(\vec r)\;\hat\psi(0, \vec x) - \vec\partial_x\hat\psi(0, \vec x)
\nonumber \\ =&\  0\,, \label{22}
\end{align}
as it should. Moreover, for an arbitrary Fock state $|\Psi\rangle$,
\begin{align}
\vec \nabla_R\,|\Psi\rangle = -\int d^3r\, \hat \psi^+(\vec r)\,\vec \partial_r\,\hat \psi(\vec r)|\Psi\rangle\,, \label{23}
\end{align}
and this holds in particular for the vacuum state $|0\rangle$.

A crucial observation was made in Ref. \cite{hfs_rec}, that every occurrence of $\vec P-q\,\vec A$ in the Coulomb gauge
can be replaced by $-q\,\vec A$ in the temporal gauge, where
the photon propagator becomes
\begin{align}
G_T^{ij}(\omega, \vec k) =&\ \frac{\rho_C(-k^2)}{k^2}\,\biggl(\delta^{ij}-\frac{k^i\,k^j}{\omega^2}\biggr)\,. \label{24}
\end{align}
The temporal  gauge is a particular case of the axial gauge and is defined by the condition
$G^{0\mu}_T = 0$ for every $\mu$. The singularity at $\omega=0$ is taken care of, as explained below.
This crucial observation leads to a significant simplification of matrix elements for all recoil corrections,
because one can use standard perturbation theory, such as a two-time Green's function approach \cite{shabaevPR},
and thus avoid cumbersome  $\vec\nabla_R$ differentiation.
Below, we present formulas for the leading two terms in the large nuclear mass expansion for hydrogenic atoms.

 \section{Pure recoil correction}
 The leading pure recoil correction is $E_\mathrm{rec} = E^{(1,0)}(Z\,\alpha)$. It is given by the first term in Eq. (\ref{17}).
 As we have already mentioned,
for the point nucleus it was first derived by Shabaev in Refs. \cite{shabaev:85,shabaev:88},
and the generalization for the finite size nucleus was achieved recently in Ref. \cite{fsrec,hfs_rec}, namely
\begin{align}
E_\mathrm{rec} =&\ \langle\bar \phi|\Sigma_\mathrm{rec}(E_D) |\phi\rangle \label{25}\\ \nonumber \\
\Sigma_\mathrm{rec}(E) =&\
\frac{i}{M} \int_s \frac{d\omega}{2\,\pi}\,  D_T^j(\omega) \,S_F(E + \omega)\,D_T^j(\omega) \,, \label{26}
\end{align}
where
\begin{align}
D_T^j(\omega,\vec r) =&\ -4\pi Z\alpha \, \gamma^i \, G_{T}^{ij}(\omega,\vec{r})\,. \label{27}
\end{align}
The subscript $s$ in the integration denotes a symmetric  integration path.
Namely, we perform Wick rotation and symmetrically integrate around the pole at $\omega=0$.
The apparent singularity at $\omega=0$ is a spurious one, as can be seen by changing to the
Coulomb gauge propagators
\begin{align}
D_C^j(\omega) =&\  D_T^j(\omega)  + \frac{1}{\omega^2}\,\big[\omega+E_D-H_D\,,\,p^j(V_C)\big]. \label{28}
\end{align}
Moreover, the temporal gauge propagator for a point nucleus  with $w = \sqrt{-\omega^2 +i\,\varepsilon}$
\begin{align}
G^{ij}_T(\omega,\vec r) =&\
-\bigg(\delta^{ij}+\frac{\nabla^i\,\nabla^j}{\omega^2}\bigg)\,\frac{e^{-w\,r}}{4\,\pi\,r} \label{29}
\end{align}
contains the Dirac $\delta$ function; therefore,
the Coulomb gauge propagators are more convenient for numerical calculations.

\section{Radiative recoil correction}
The exact, in $Z\,\alpha$, radiative recoil correction $E_\mathrm{radrec} = E^{(1,1)}$ has not yet been published in the literature.
We split it into self-energy and vacuum-polarization parts
\begin{align}
E_\mathrm{radrec} = E_\mathrm{selfrec} + E_\mathrm{vprec}\,. \label{30}
\end{align}
The vacuum polarization part can be implemented by the effective charge density $ \rho_\mathrm{vp}(-k^2)$, namely using Eqs. (\ref{09}) and (\ref{10}),
\begin{align}
\rho_\mathrm{Cvp}(-k^2) =   -\bar\omega(k^2)\,\rho_C(-k^2)\,, \label{31}
\end{align}
and thus takes the form analogous to Eq. (\ref{26})
\begin{align}
E_\mathrm{vprec} =&\ \delta_\mathrm{vp}
\frac{i}{M} \int_s \frac{d\omega}{2\,\pi}\, \lbr \bar\phi | D_T^j(\omega) \,S_F(E_D + \omega)\,D_T^j(\omega) | \phi \rbr, \label{32}
\end{align}
where $\delta_\mathrm{vp}$ perturbs $\phi, H_D, E_D$, and $D^j_T$, whenever $\rho_C$ is present, namely
\begin{align}
E_\mathrm{vprec} =&\
\frac{i}{M} \int_s \frac{d\omega}{2\,\pi}\, \Bigl[
2\,\lbr \bar\phi | D_{T\mathrm{vp}}^j(\omega) \,S_F(E_D + \omega)\,D_T^j(\omega) | \phi \rbr
\nonumber \\ &\hspace*{-3ex}
+2\,\lbr \bar\phi | \gamma^0\,V_\mathrm{Cvp}\,S'_F(E_D)\,D_{T}^j(\omega) \,G(E_D + \omega)\,D_T^j(\omega) | \phi \rbr
\nonumber \\ &\hspace*{-3ex}
+\lbr \bar\phi | D_{T}^j(\omega) \,S_F(E_D + \omega)\,\gamma^0\,(V_\mathrm{Cvp} - \langle V_\mathrm{Cvp}\rangle)
\nonumber \\ &\hspace*{-3ex}
\times S_F(E_D + \omega)\,D_T^j(\omega) | \phi \rbr \Bigr]\,,
\label{33}
\end{align}
and where $S'_F(E_D)$ is the reduced Dirac propagator (the reference state with the energy $E_D$ is subtracted out).

The self-energy part is obtained as follows. The fermion propagator
\begin{align}
\frac{1}{\not\!p-m-\Sigma(E)} \label{34}
\end{align}
has corrections due to the self-energy and the recoil
\begin{align}
\Sigma(E)= \Sigma_\mathrm{rad}(E) + \Sigma_\mathrm{rec}(E) + \Sigma_\mathrm{radrec}(E) + \ldots\,, \label{35}
\end{align}
where $\Sigma_\mathrm{rad}$ is defined in Eq. (\ref{07}), $\Sigma_\mathrm{rec}$ in Eq. (\ref{26}), and
\begin{widetext}
\begin{align}
\Sigma_\mathrm{radrec}(E) =&\
\frac{i}{M} \int_s \frac{d\omega'}{2\,\pi}\,  e^2\!\int\!\frac{d^4k}{(2\,\pi)^4\,i}\,\frac{1}{k^2}
\nonumber \\ \times&\ \Bigl[
\gamma^\mu\,e^{-i\,\vec k\cdot\vec r}\,S_F(E+\omega)\,D_T^j(\omega') \,S_F(E + \omega+\omega')\,D_T^j(\omega')\,S_F(E+\omega)\,\gamma_\mu\,e^{i\,\vec k\cdot\vec r}
\nonumber \\ &\ +
D_T^j(\omega') \,S_F(E + \omega') \, \gamma^\mu\,e^{-i\,\vec k\cdot\vec r}\,S_F(E+\omega + \omega')\,\gamma_\mu\,e^{i\,\vec k\cdot\vec r}\, S_F(E + \omega') \,D_T^j(\omega')
\nonumber \\ &\ +
\gamma^\mu\,e^{-i\,\vec k\cdot\vec r}\,S_F(E+\omega)\,D_T^j(\omega') \,S_F(E + \omega+\omega')\,\,\gamma_\mu\,e^{i\,\vec k\cdot\vec r}\, S_F(E + \omega') \,D_T^j(\omega')
\nonumber \\ &\ +
D_T^j(\omega') \,S_F(E + \omega') \, \gamma^\mu\,e^{-i\,\vec k\cdot\vec r}\,S_F(E+\omega + \omega')\,\,D_T^j(\omega')\,S_F(E+\omega)\,\gamma_\mu\,e^{i\,\vec k\cdot\vec r}
\Bigr]
\label{36}
\end{align}
\end{widetext}
is a sum of all one-particle irreducible diagrams (in the temporal gauge).
The change in the position of the pole of the fermion propagator in Eq. (\ref{34}) due to the presence of $\Sigma(E)$ is
\begin{align}
E_\mathrm{selfrec} =&\ \langle\bar\phi|\Sigma_\mathrm{radrec}(E_D) |\phi\rangle
\nonumber \\ &
+ 2\,\langle\bar\phi| \Sigma_\mathrm{rad}(E_D)\,S'_F(E_D)\, \Sigma_\mathrm{rec}(E_D) |\phi\rangle
\nonumber \\ &
 +\langle\bar\phi|\Sigma'_\mathrm{rad}(E_D) |\phi\rangle\,\langle\bar\phi|\Sigma_\mathrm{rec}(E_D) |\phi\rangle
 \nonumber \\ &
 +\langle\bar\phi|\Sigma'_\mathrm{rec}(E_D) |\phi\rangle\,\langle\bar\phi|\Sigma_\mathrm{rad}(E_D) |\phi\rangle\,.
 \label{37}
\end{align}
This radiative recoil correction has been calculated only up to $\alpha\,(Z\,\alpha)^5$ order,
and the higher-order terms are unknown, which limits the accuracy of the hydrogen Lamb shift \cite{codata}. Moreover, we expect this correction
to be significant for light muonic atoms, where the electron vacuum-polarization is combined with a relatively large
muon-nucleus mass ratio. Therefore, we perform exemplary calculation of this correction for several hydrogenic ions in Sec. VIII.

\section{Recoil correction to the hyperfine splitting}
The nonperturbative formula for the  recoil correction to the hyperfine splitting in hydrogen-like ions has recently been derived in Ref. \cite{hfs_rec}.
It comes from the NRQED Hamiltonian of the nucleus in Eq. (\ref{15}) ( including the relevant terms only)
 \begin{align}
H_\mathrm{nuc} =&\ \frac{\vec\Pi^2}{2\,M} -\frac{q}{2\,M}\,g\,\vec I\cdot\vec B
\nonumber \\ &\
-\frac{q}{4\,M^2}\,(g-1)\,\vec I\cdot[\vec E\times\vec\Pi-\vec\Pi\times\vec E]\,,
\label{38}
\end{align}
and takes the form
\begin{align}
E_\mathrm{hfsrec} =&\ E_\mathrm{kin} + E_\mathrm{so} + E_\mathrm{sec}\,, \label{39}
\end{align}
where
\begin{align}
E_\mathrm{kin} = &\
\frac{1}{M}\! \int_s \frac{d\omega}{2\,\pi}\,\frac{1}{\omega}
\bigl[ \lbr \bar\phi | D_T^j(\omega) \,S_F(E_D+\omega)\, \partial^j(V_\mathrm{hfs}(\omega)) | \phi \rbr
\nonumber \\ &\
-  \lbr \bar\phi | \partial^j(V_\mathrm{hfs}(\omega)) \,S_F(E_D+\omega)\, D_T^j(\omega) | \phi \rbr \bigr]
\nonumber \\ &\
+ \delta_\mathrm{hfs}\frac{i}{M} \int_s \frac{d\omega}{2\,\pi}\, \lbr \bar\phi | D_T^j(\omega)\,S_F(E_D+\omega )\, D_T^j(\omega) | \phi \rbr\,,
\label{40}\\
E_\mathrm{so} = &\
-\frac{(g-1)}{M^2}\,\epsilon^{ijk}\,I^i \int_s \frac{d\,\omega}{2\,\pi}\,\omega
\nonumber \\ & \times
\langle \bar\phi | D_T^j(\omega)\, S_F(E_D+\omega)\,D_T^k(\omega) | \phi \rangle\,,  \label{41} \\ \nonumber \\
E_\mathrm{sec} = &\
\biggl(\frac{4\,\pi\,Z\,\alpha}{2\,M}\,g\biggr)^2\,  \epsilon^{ijk}\,I^k  \int_s \frac{d\,\omega}{2\,\pi}\,\frac{1}{\omega}
\nonumber \\ & \hspace*{-3ex} \times
\langle \bar\phi |(\vec\gamma\times\vec \nabla)^i\,D(\omega)\,
S_F(E_D+\omega)\,(\vec\gamma\times\vec \nabla)^j\,D(\omega)\,| \phi \rangle\,, \label{42}
\end{align}
where
\begin{align}
{D}(\omega,r) =&\ \int \frac{d^3k}{(2\pi)^3}\, e^{i\vec{k}\cdot\vec{r}}\,\frac{\rho({\vec k}^2-\omega^2)}{\omega^2-{\vec k}^2}
\label{43}
\end{align}
and
\begin{align}
V_\mathrm{hfs}(\omega,\vec r) =&\  e\,\vec \mu\cdot \vec \gamma\times\vec\nabla D(\omega, r)\,,
\label{44}
\end{align}
such that $V_\mathrm{hfs}(0,r) = V_\mathrm{hfs}(r)$, where
\begin{align}
V_\mathrm{hfs}(r) =&\ \frac{e}{4\,\pi}\,\vec \mu\cdot\vec\gamma\times\biggl[\frac{\vec r}{r^3}\biggr]_\mathrm{fs}\,.
\label{45}
\end{align}
This recoil correction has not yet been numerically calculated. Expansion terms up to $(Z\,\alpha)^2\,E_F$ are presently known \cite{Bodwin:88},
but the higher-order recoil terms have not yet been studied.

\section{Second-order recoil correction}
The second-order recoil correction $ E_\mathrm{quadrec} = E^{(2,0)}$ has also not yet been considered in the literature.
It comes from the NRQED Hamiltonian for the nucleus in Eq. (\ref{15}) (including relevant terms only)
\begin{align}
H_\mathrm{nuc} =&\ \frac{\vec\Pi^2}{2\,M} -\frac{q}{2\,M}\,g\,\vec I\cdot\vec B - \frac{q\,\delta_I}{8\,M^2}\,\vec\nabla\cdot\vec E\,,
\label{46}
\end{align}
and is split into three parts,
 \begin{align}
 E_\mathrm{quadrec} = E_\mathrm{kin} +  E_\mathrm{mag} + E_\mathrm{zit}\,, \label{47}
 \end{align}
 which are obtained as follows.

The kinetic energy part $E_\mathrm{kin}$ is the second-order correction due to the nonrelativistic kinetic energy of the nucleus
\begin{align}
E_\mathrm{kin} =&\ \langle\bar\phi|\Sigma_\mathrm{quadrec}(E_D) |\phi\rangle
\nonumber \\ &
+ \langle\bar\phi| \Sigma_\mathrm{rec}(E_D)\,S'(E_D)\, \Sigma_\mathrm{rec}(E_D) |\phi\rangle
\nonumber \\ &
 +\langle\bar\phi|\Sigma'_\mathrm{rec}(E_D) |\phi\rangle\,\langle\bar\phi|\Sigma_\mathrm{rec}(E_D) |\phi\rangle\,,
 \label{48}
\end{align}
where $\Sigma_\mathrm{rec}(E)$ is defined in Eq. (\ref{26}), and $\Sigma_\mathrm{quadrec}(E)$ is a sum of all
one-particle irreducible diagrams, namely
\begin{widetext}
\begin{align}
\Sigma_\mathrm{quadrec}(E) =&\ \bigg(\frac{i}{M} \int_s \frac{d\omega_1}{2\,\pi}\bigg)\,\bigg(\frac{i}{M} \int_s \frac{d\omega_2}{2\,\pi}\bigg)
\nonumber \\ &\
\Bigl[
D_T^j(\omega_1) \,S_F(E_D + \omega_1)\,D_T^j(\omega_2)\,S_F(E_D + \omega_1+\omega_2)\,D_T^j(\omega_1)\,S_F(E_D + \omega_2)\,D_T^j(\omega_2)
\nonumber \\ &\ +
D_T^j(\omega_1) \,S_F(E_D + \omega_1)\,D_T^j(\omega_2)\,S_F(E_D + \omega_1+\omega_2)\,D_T^j(\omega_2)\,S_F(E_D + \omega_1)\,D_T^j(\omega_1)
\Bigr]
\label{49}
\end{align}
\end{widetext}

The magnetic contribution $E_\mathrm{mag}$ is the second-order correction due to $-\vec \mu\cdot\vec B$ in Eq. (\ref{46})
\begin{align}
E_{\rm mag} =&\ \langle \bar\phi| \gamma^0\,V_\mathrm{hfs}\,S'_F(E_D)\, \gamma^0\,V_\mathrm{hfs} |\phi\rangle\,, \label{50}
\end{align}
and the product of the nuclear magnetic moments in the above is to be replaced by
$ \mu^a\,\mu^b \rightarrow  \delta^{ab}\,\vec\mu^{\,2}/3$. We note that the antisymmetric part would contribute
to the hyperfine splitting, but it would be incorrect. The correct formula was presented in the previous section in Eq. (\ref{42}).

Finally, the {\em zitterbewegung} term $E_\mathrm{zit}$ depends on the nuclear spin. It vanishes for scalar particles (nuclei),
while for spin $1/2$ it is $\delta_{1/2} = 1$, and
\begin{align}
E_\mathrm{zit} =&\   \frac{\delta_I}{8\,M^2}\,\langle \vec\nabla^2(V_C) \rangle \label{51}
\end{align}
The numerical calculation of this second-order correction is not trivial, but what is important is that
all such formulas can be written down.

\section{Numerical calculation of the nuclear recoil electron vacuum-polarization correction in muonic atoms}
\begin{table*}[t]
    \caption{
    Nuclear-recoil--vacuum-polarization correction to the
    $2P_{1/2}$-$2S$ transition energy of H-like muonic ions, in meV.
    \label{tab:vprec}}
\begin{ruledtabular}
    \begin{tabular}{l w{4.7} w{4.7} w{4.7} w{5.5} w{6.4} w{6.4} }
    & \multicolumn{1}{c}{$^1$H}
    & \multicolumn{1}{c}{$^4$He}
    & \multicolumn{1}{c}{$^7$Li}
    & \multicolumn{1}{c}{$^{12}$C}
    & \multicolumn{1}{c}{$^{132}$Xe}
    & \multicolumn{1}{c}{$^{208}$Pb}\\
\hline\\[-5pt]
$M/m_{\mu}$	& \multicolumn{1}{c}{$8.880243$}
           &    \multicolumn{1}{c}{$35.27765$}
                	&	\multicolumn{1}{c}{$61.83924$}
                       &  \multicolumn{1}{c}{$105.7641$}
                            &  \multicolumn{1}{c}{$1162.618$}
                                &  \multicolumn{1}{c}{$1833.145$} \\
$r_C$ [fm]
& \multicolumn{1}{c}{$0.84060$}
           &    \multicolumn{1}{c}{$1.67860$}
                	&	\multicolumn{1}{c}{$2.4440$}
                       &  \multicolumn{1}{c}{$2.4702$}
                            &  \multicolumn{1}{c}{$4.7859$}
                                &  \multicolumn{1}{c}{$5.5012$} \\
\hline\\[-5pt]
$E_{\rm vprec, el}$
	    & -56.36324	 &    -78.41878	&   -107.1656  &   -241.863	&	5407.20  &	6895.5 \\
$E_{\rm vprec, ph}$
	    &   0.00555  &      0.03214 &      0.1068  &      1.191 &   -293.61  &  -971.2 \\
$E_{\rm vprec}$
        & -56.35769   &   -78.38664  &  -107.0588  &   -240.672	&	5113.59  &  5924.3 \\
$E_{\rm vprec}({\rm l.o.})$
        & -56.36185   &   -78.60251  &  -108.66    &   -258.91  &  -1825.  & -2666. \\
    \end{tabular}
\end{ruledtabular}
\end{table*}

This is an exemplary numerical calculation of the nuclear-recoil electron vacuum-polarization (EVP) correction for several
muonic (hydrogenlike) atoms. This correction, given by Eq.~(\ref{32}), we transform to the Coulomb gauge
\begin{align}
E_{\rm vprec} =
 \frac{m_\mu^2}{M}\frac{i}{2\pi}  & \int_{-\infty}^{\infty} d\omega\,
\delta_\mathrm{vp} \lbr \phi | \big[ p^j - D_C^j(\omega)\big] \,
  \nonumber \\ & \times
 [E_D-H_D]^{-1} \, \big[ p^j - D_C^j(\omega)\big] | \phi \rbr\,, \label{52}
\end{align}
where  $m_{\mu}$ is the muon mass, $\delta_{\rm vp}$ denotes perturbation by the electron vacuum polarization, 
$D_C^j(\omega) = -4\pi\, Z\alpha \, \alpha^i \, G_{C}^{ij}(\omega,\vec{r})$
and $G_C^{ij}$ is the transverse part of the photon propagator in the Coulomb gauge given by
\begin{align}
G_{C}^{ij}(\omega,\vec{r}) = \delta^{ij}\,{D}(\omega,r) + \frac{\nabla^i\nabla^j}{\omega^2} \, \Big[ {D}(\omega,r)
 - {D}(0,r)\Big]\,, \label{53}
\end{align}
and
\begin{align}
D(\omega,r) = \int \frac{d^3k}{(2\pi)^3}\, e^{i\vec{k}\cdot\vec{r}}\,\frac{\rho({\vec k}^2-\omega^2)}{\omega^2-{\vec k}^2}\,.
\end{align}
See Ref.~\cite{fsrec} for details.

We divide $E_{\rm vprec}$ into two parts,
\begin{align}
E_{\rm vprec} = E_{\rm vprec, ph} + E_{\rm vprec, el}\,, \label{54}
\end{align}
where $E_{\rm vprec, ph}$ is due to perturbation of $D_C(\omega)$ by vacuum polarization,
whereas $E_{\rm vprec, el}$
is the remainder. The calculation of $E_{\rm vprec, el}$ is relatively straightforward. We include the
vacuum-polarization potential $V_{\rm Cvp}(r)$ into the Dirac Hamiltonian, calculate the
recoil correction using the procedure developed in Ref.~\cite{fsrec}, and then linearize
with respect to $V_{\rm Cvp}$ by taking the numerical derivative.

The computation of $E_{\rm vprec, ph}$ is more complicated. We write it in the Coulomb gauge as
follows,
\begin{align}
E_{\rm vprec, ph} = &\,
     \frac{m_\mu^2}{M}\,\frac{i}{\pi} \int_{-\infty}^{\infty}d\omega\,
\sum_n
 \frac1{E_D + \omega - E_n(1-i0)}\,
\nonumber \\ & \times
 \Big[
 -\bra{\phi} \vec{p}\ket{n}\,
   \bra{n}\delta_\mathrm{vp}\vec{D}_C(\omega) \ket{\phi}\,,
   \nonumber \\ &
+
  \bra{\phi} \vec{D}_C(\omega)\ket{n} \,
   \bra{n} \delta_\mathrm{vp}\vec{D}_C(\omega) \ket{\phi}
   \Big]
   \,. \label{55}
\end{align}
The function $\delta_\mathrm{vp}D_C^j(\omega)$ is obtained from $D_C^j(\omega)$ by modifying the photon
propagator with vacuum polarization. This modification leads to the replacement
\begin{align}
{D}(\omega,r) \to {D}_{\rm vp}(\omega,r)\,, \label{56}
\end{align}
where
\begin{align}
{D}_{\rm vp}(\omega,r) =&\
\int \frac{d^3k}{(2\pi)^3}\, e^{i\vec{k}\cdot\vec{r}}\,\frac{\rho({\vec k}^2-\omega^2)}{\omega^2-{\vec k}^2}\,[-\bar\omega(\omega^2-\vec k^{\,2})]\,,
\label{57}
\end{align}
and $\bar{\omega}(k^2)$ is defined in Eq. (\ref{10}).
Let us now obtain ${D}_{\rm vp}$ for the exponential model of the nuclear charge
distribution
\begin{align}
\rho(\vec k^2) =&\ \frac{\lambda^4}{(\lambda^2 + \vec k^{\,2})^2}\,. \label{58}
\end{align}
We first consider the case of pure imaginary $\omega$. For $w^2 = -\omega^2$, we get
\begin{align}
{D}_{\rm vp}(i\,w,r) =&\
-\frac{\alpha}{\pi} \int_4^\infty d(q^2) \frac{u(q^2)}{q^2}\int \frac{d^3k}{(2\pi)^3}\, e^{i\vec{k}\cdot\vec{r}}\,X \,, \label{59}
\end{align}
where
\begin{align}
X =&\
 \frac{1}{(m_e^2\,q^2 +w^2 + \vec k^{\,2})}\,\frac{\lambda^4}{(\lambda^2 + w^2 + \vec k^{\,2})^2}
 \nonumber \\ =&\
 \frac{\lambda^4}{(\lambda^2-m_e^2\,q^2)^2}\,\biggl[
 \frac{1}{m_e^2\,q^2 + w^2+ \vec k^{\,2}}
 \nonumber \\ &\ -\frac{1}{\lambda^2+w^2+\vec k^{\,2}} - \frac{\lambda^2-m_e^2\,q^2}
 {(\lambda^2+w^2+\vec k^{\,2})^2}\biggr]\,.
 \label{60}
\end{align}
Performing integration over $\vec k$, we obtain
\begin{align}
{D}_{\rm vp}(i\,w,r) =&\
-\frac1{4\pi}
\int_2^{\infty}dq\, A_{\rm vp}(q)\,
  \bigg[ \frac{e^{-\sqrt{w^2+m_e^2q^2}\,r}}{r}
\nonumber \\ &
  - \frac{e^{-\sqrt{w^2+\lambda^2}\,r}}{r}
  - \frac{\lambda^2-m_e^2q^2}{2\sqrt{\omega^2+\lambda^2}}\,
   e^{-\sqrt{w^2+\lambda^2}\,r}
 \bigg]
 \,,  \label{61}
\end{align}
where
\begin{align}
A_{\rm vp}(q) = \frac{\alpha}{\pi}\,\frac23\,\frac{\sqrt{q^2-4}\,(2+q^2)}{q^4}
\frac1{(1-m_e^2q^2/\lambda^2)^2}\,.  \label{62}
\end{align}
For the general complex $\omega$, we analytically continue the above formulas and obtain
\begin{align}
{D}_{\rm vp}(\omega,r) =&\
-\frac1{4\pi}
\int_2^{\infty}dq\, A_{\rm vp}(q)\,
  \bigg[ \frac{e^{i\sqrt{\omega^2-m_e^2q^2}\,r}}{r}
\nonumber \\ &
  - \frac{e^{i\sqrt{\omega^2-\lambda^2}\,r}}{r}
  - \frac{i(\lambda^2-m_e^2q^2)}{2\sqrt{\omega^2-\lambda^2}}\,
   e^{i\sqrt{\omega^2-\lambda^2}\,r}
 \bigg]
 \,.  \label{63}
\end{align}
We note that for $\omega = 0$, the function ${D}_{\rm vp}$ becomes proportional to the
vacuum-polarization potential $V_{\rm Cvp}$,
\begin{align}
V_{\rm Cvp}(r) = 4\pi\Za\,{D}_{\rm vp}(0,r) \,.  \label{64}
\end{align}

Results of our numerical calculations of the recoil-vacuum-polarization correction
are presented in Table~\ref{tab:vprec} for the
$2P_{1/2}$-$2S$ transition energy of
several muonic H-like ions, together
with the contribution of the leading order of the $\Za$ expansion of this correction,
\begin{align}
E_{\rm vprec}({\rm l.o.}) = &\ -\frac{2}{3\pi}\,\frac{m_{\mu}^2c^2}{M}\,
\alpha(\Za)^2
 \nonumber \\&
  \int_1^{\infty}du\,
  \frac{\beta^2 (-1 + 6\beta u)\sqrt{u^2-1} (1 + 2 u^2)}
  {u^2 (1 + 2\beta u)^5}\,,  \label{65}
\end{align}
where $\beta = m_e/(\Za\, m_{\mu})$.

It can be seen that the deviation of the all-order result from the lowest-order term
is quite small for hydrogen and helium (as expected), but grows fast with the
increase of the nuclear charge. For heavy ions, the lowest-order formula
does not reproduce even the overall sign of the total correction.
Moreover, we note that for light ions the nuclear-recoil electron vacuum-polarization correction
is larger than the pure nuclear recoil for this particular transition.  So, for muonic hydrogen, the
first-order in $m/M$ nuclear recoil contributes 1.63~meV to the $2P_{1/2}$-$2S$
energy difference, whereas the nuclear-recoil with vacuum-polarization contributes
$-56.4$~meV.

\section{Summary}

HPQED enables the systematic inclusion of finite nuclear mass corrections using exact formulas in terms of $Z\alpha$. By recognizing that the nuclear momentum $\vec{P} - q\vec{A}$ can be represented as $-q\vec{A}$ in the temporal gauge, the derivation falls within the realm of standard bound state perturbative QED and can thus be performed straightforwardly. The
obtained formulas can be extended to many-electron systems using the $1/Z$ expansion \cite{shabaevPR}, similarly to the leading recoil correction \cite{few_recoil}. Numerical calculations would likely require transforming to the Coulomb gauge, resulting in much longer formulas. For electronic atoms,
most of higher-order recoil corrections derived here are probably insignificant, except for the radiative recoil, which is required
for the improved description of the hydrogen Lamb shift.
However, for muonic atoms, these corrections could lead to substantial effects since muons are 206 times heavier than electrons.
In fact, there is an ongoing  QUARTET project \cite{ohayon} to determine nuclear charge radii from the spectra of $3\leq Z\leq 10$ muonic atoms.
An additional benefit of the obtained formulas is the complete inclusion of finite nuclear size (but not the nuclear polarizability,
which needs to be accounted for separately).
Finally, the derived formulas can be used to obtain higher-order recoil corrections in the $Z\alpha$ expansion, such as $\alpha (Z\alpha)^6 m/M$, which have yet to be addressed in the literature.

\acknowledgments
KP wishes to thank Frederic Merkt for his hospitality during his stay at ETH, Zurich,
where this paper has been written.

\end{document}